\documentclass[twocolumn,showpacs,preprintnumbers,amsmath,amssymb,floatfix]{revtex4}

\usepackage{graphicx}
\usepackage{epsfig}
\usepackage{amsfonts}
\usepackage{amssymb}
\usepackage{epsf}
\usepackage{color}

\usepackage{tabularx} 
\usepackage{amsmath}  
\usepackage[final]{hyperref} 
\hypersetup{
	colorlinks=true,       
	linkcolor=blue,        
	citecolor=blue,        
	filecolor=magenta,     
	urlcolor=blue         
	}

\def\beq{\begin{equation}}
\def\eeq{\end{equation}}
\def\bea{\begin{eqnarray}}
\def\eea{\end{eqnarray}}

\begin{document}


\title{Evasion of No-Hair Theorems and Novel Black-Hole Solutions in Gauss-Bonnet Theories}


\author{{\bf G. Antoniou}, {\bf A. Bakopoulos} and {\bf P. Kanti}}

\affiliation{Department of Physics,
University of Ioannina, Ioannina GR-45110, Greece}

\bigskip \medskip
\begin{abstract}
We consider a general Einstein-scalar-GB theory with a coupling function
$f(\phi)$. We demonstrate that black-hole solutions appear as a generic feature of
this theory since a regular horizon and an asymptotically-flat solution may be easily
constructed under mild assumptions for $f(\phi)$. We show that the existing no-hair
theorems are easily evaded, and a large number of regular, black-hole solutions with
scalar hair are then presented for a plethora of coupling functions $f(\phi)$.
\end{abstract}

\maketitle


\setcounter{page}{1}

\section{Introduction}

The existence or not of black holes associated with a non-trivial scalar field in the exterior
region has attracted the attention of researchers over a period of many decades. Early on,
a {\it no-hair theorem} \cite{NH-scalar} appeared, that excluded static black holes with a
scalar field, but this was soon outdated by the discovery of black holes with Yang-Mills
\cite{YM} or Skyrme fields \cite{Skyrmions}. The emergence of additional solutions where
the scalar field had a conformal coupling to gravity \cite{Conformal} led to the
formulation of a novel no-hair theorem \cite{Bekenstein} (for a review, see 
\cite{Herdeiro}). Recently, this argument
was extended to the case of standard scalar-tensor theories \cite{SF}, and a new
form was proposed, that covers the case of Galileon fields \cite{HN}. 

However, both novel forms of the no-hair theorem \cite{Bekenstein, HN} were shown
to be evaded: the former in the context of the Einstein-Dilaton-Gauss-Bonnet theory
\cite{DBH} and the latter in a special case of shift-symmetric Galileon theories
\cite{Babichev, SZ, Benkel}.
Common feature of the above theories was the presence
of the quadratic Gauss-Bonnet (GB) term defined as
$R^2_{GB}=R_{\mu\nu\rho\sigma}R^{\mu\nu\rho\sigma}-4R_{\mu\nu}R^{\mu\nu}+R^2$,
in terms of the Riemann tensor $R_{\mu\nu\rho\sigma}$, the Ricci tensor $R_{\mu\nu}$
and the Ricci scalar $R$. In both cases, basic requirements of the no-hair theorems
were invalidated, and this paved the way for the construction of
the counter-examples. 

Here, we consider a general class of scalar-GB theories,
of which the cases \cite{DBH, SZ} constitute particular examples. We demonstrate
that black-hole solutions, with a regular horizon and an asymptotically-flat limit
may in fact be constructed for a large class of such theories under mild only
constraints on the coupling function $f(\phi)$ between the scalar field and the GB term.
We address the requirements of both the old and novel no-hair theorems, and we
show that they are not applicable for this specific class of theories. In accordance to the
above,
we then present a large number of exact, regular black-hole solutions with scalar hair
for a variety of forms for the coupling function.


\section{The Einstein-Scalar-GB Theory}

We first consider the following action functional
\begin{equation}\label{action}
S=\frac{1}{16\pi}\int{d^4x \sqrt{-g}\left[R-\frac{1}{2}\,\partial_{\mu}\phi\,\partial^{\mu}\phi+f(\phi)R^2_{GB}\right]},
\end{equation}
that describes a  generalised gravitational theory containing the Ricci
scalar curvature $R$, a scalar field $\phi$, and the quadratic Gauss-Bonnet
term $R^2_{GB}$. The latter, being a total derivative in four dimensions,
is coupled to $\phi$ through a coupling function $f(\phi)$. The form of the
latter quantity may be inspired from either string-theory \cite{Metsaev}
or Horndeski theory \cite{Horndeski}. Note that, in
this work, we employ units in which $G=c=1$.

By varying the action (\ref{action}) with respect to the metric tensor $g_{\mu\nu}$
and the scalar field $\phi$, we derive the gravitational field equations and the
equation for the scalar field, respectively. These have the covariant form:
\begin{eqnarray}
& G_{\mu\nu}=T_{\mu\nu}\,,& \label{Einstein-eqs}\\[1mm]
&\nabla^2 \phi+\dot{f}(\phi)R^2_{GB}=0\,,& \label{scalar-eq}
\end{eqnarray}
where $G_{\mu\nu}$ is the Einstein tensor, and a dot
denotes the derivative with respect to the scalar field. Also, 
\begin{eqnarray}\label{f}
T_{\mu\nu}&=&-\frac{1}{4}g_{\mu\nu}\partial_{\rho}\phi\partial^{\rho}\phi+\frac{1}{2}\partial_{\mu}\phi\partial_{\nu}\phi \nonumber \\[0mm]
&-& 
\frac{1}{2}(g_{\rho\mu}g_{\lambda\nu}+g_{\lambda\mu}g_{\rho\nu})
\eta^{\kappa\lambda\alpha\beta}\tilde{R}^{\rho\gamma}_{\quad\alpha\beta}
\nabla_{\gamma}\partial_{\kappa}f,
\end{eqnarray}
where $\tilde{R}^{\rho\gamma}_{\quad\alpha\beta}=\eta^{\rho\gamma\sigma\tau}
R_{\sigma\tau\alpha\beta}=\epsilon^{\rho\gamma\sigma\tau}
R_{\sigma\tau\alpha\beta}/\sqrt{-g}$. Note that the energy-momentum tensor
$T_{\mu\nu}$ receives contributions from both the scalar field and the
Gauss-Bonnet term.

In the context of the above theory, we will seek spherically-symmetric solutions,
with a line-element
\begin{equation}\label{metric}
{ds}^2=-e^{A(r)}{dt}^2+e^{B(r)}{dr}^2+r^2({d\theta}^2+\sin^2\theta\,{d\varphi}^2)\,,
\end{equation}
that describe regular, static, asymptotically-flat black holes. Our analysis
will also investigate the general constraints that the coupling function
$f(\phi)$ needs to obey in order for these solutions to arise. 

By employing the line-element (\ref{metric}), the Einstein's equations take
the explicit form
\bea
4e^B(e^{B}+rB'-1)&=&\phi'^2\bigl[r^2e^B+16\ddot{f}(e^B-1)\bigr] \nonumber\\
&& \hspace*{-2.5cm}-8\dot{f}\left[B'\phi'(e^B-3)-2\phi''(e^B-1)\right], \label{tt-eq}
\eea
\vskip -0.6cm
\beq
4e^B(e^B- r A'-1)=-\phi'^2 r^2 e^B +8\left(e^B-3\right)\dot{f}A'\phi',
\label{rr-eq}
\eeq
\vskip -0.6cm
\begin{eqnarray}
&&\hspace*{-0.7cm} e^B\bigl[r{A'}^2-2B'+A'(2-rB')+2rA''\bigr]= -\phi'^2 r e^B
\nonumber \\
&&\hspace*{-0.8cm} +8 \phi'^2 \ddot{f}A'+ 
4\dot{f}[\phi'(A'^2+2A'')+A'(2\phi''-3B'\phi')], \label{thth-eq}
\end{eqnarray}
while the scalar equation reads:
\bea
&& \hspace*{-0.3cm}2r\phi''+(4+rA'-rB')\,\phi'+
\frac{4\dot{f}e^{-B}}{r}\bigl[(e^B-3)A'B'\nonumber \\
&& \hspace*{1.5cm}-(e^B-1)(2A''+A'^2)\bigr]=0\,. \label{phi-eq}
\eea
In the above, the prime denotes differentiation with respect to $r$ -- throughout 
this work, we assume that the scalar field shares the symmetries of the spacetime.

Equation (\ref{rr-eq}) may take the form of a second-order polynomial
with respect to $e^B$, which can then be solved to give 
$e^B=(-\beta\pm\sqrt{\beta^2-4\gamma})/2$,
where
\beq
\beta=\frac{r^2{\phi'}^2}{4}-(2\dot{f}\phi'+r) A'-1\,, \qquad 
\gamma=6\dot{f}\phi'A'\label{bg}\,.
\eeq
Then, eliminating $B$ from the set of the remaining equations (\ref{tt-eq}),
(\ref{thth-eq}) and (\ref{phi-eq}), we may form a system of two independent,
ordinary differential equations of second order for the functions $A$ and
$\phi$:
\beq
A''=\frac{P}{S}\,,\qquad
\phi''=\frac{Q}{S}\,, \label{Aphi-system}
\eeq 
where the functions $P$, $Q$ and $S$ are lengthy expressions of $(r, \phi', A', \dot f,
\ddot f)$.



We will now demonstrate that our set of equations, with a general coupling function $f(\phi)$,
allows for the construction of a black-hole solution with a regular horizon  provided that
$f$ satisfies certain constraints. For a spherically-symmetric spacetime, the presence of
a horizon is realised for $e^A \rightarrow 0$, as $r \rightarrow r_h$, or equivalently for
$A' \rightarrow \infty$ -- the latter will be used in our analysis as an assumption
but it will be shown to follow from the former. On the other hand, the regularity of
the horizon amounts to demanding that
$\phi$, $\phi'$ and $\phi''$ remain finite in the limit $r \rightarrow r_h$. Then,
assuming that $A' \rightarrow \infty$ while $\phi'$ remains finite
\footnote{Note that, in the expression of $e^B$, we
have kept only the (+) sign as the (-) sign leads to $e^B \simeq {\cal O}(1)$, which
is not a black-hole solution.}, Eqs. (\ref{Aphi-system})
take the approximate forms
\bea
&&\hspace*{-1cm}A''=-\frac{r^4+4r^3\phi'\dot{f}+4r^2\phi'^2\dot{f}^2-24\dot{f}^2}{r^4+2r^3\phi'\dot{f}-48\dot{f}^2}A'^2+...\label{A-approx-h}\\
&&\hspace*{-1cm}\phi''=-\frac{(2\phi'\dot{f}+r)(r^3\phi'+12\dot{f}+2r^2\phi'^2\dot{f})}{r^4+2r^3\phi'\dot{f}-48\dot{f}^2}A'+...\label{phi-approx-h}
\eea
Focusing on the second of the above equations, we observe that $\phi''$ diverges at
the horizon if $f(\phi)$ is either zero or left unconstrained. However, $\phi''$ may
be rendered finite if either one of the two expressions in the numerator of 
Eq. (\ref{phi-approx-h}) is zero close to the horizon. 

If we assume that $(2\phi'\dot{f}+r)=0$ close to the horizon, then a careful
inspection of our equations reveals that, in that case, $\phi'' \simeq \sqrt{A'}/\dot f$.
Thus, for $\phi''$ to remain finite, we must demand that $\dot f \rightarrow \pm\infty$, 
near the horizon. This may be easily shown to lead to either a divergent or a trivial
scalar field near the horizon, for every elementary form of $f(\phi)$ we have tried. 
Therefore, for the construction of a regular horizon in the presence of a non-trivial
scalar field, we are led to consider the second choice: 
$r^3\phi'+12\dot{f}+2r^2{\phi'}^2\dot{f}=0$. This may be
easily solved to yield:
\begin{equation}\label{con-phi'}
\phi'_h=\frac{r_h}{4\dot{f}_h}\left(-1\pm\sqrt{1-\frac{96\dot{f}_h^2}{r_h^4}}\right),
\end{equation}
where all quantities have been evaluated at $r_h$.
To ensure that $\phi'_h$ is real, we must impose the following constraint on
the coupling function
\begin{equation}\label{con-f}
\dot{f}_h^2<\frac{r_h^4}{96}\,.
\end{equation}

Turning now to Eq. (\ref{A-approx-h}), and using the constraint (\ref{con-phi'}),
we find that the coefficient of $A'^2$  simplifies to $-1$. Then, upon
integration with respect to $r$, we obtain $A'=(r-r_h)^{-1}+\mathcal{O}(1)$
which, in accordance to our initial assumption, diverges close to the horizon. 
Integrating once more and putting everything together, we may write the 
near-horizon solution as
\bea
&&e^{A}=a_1 (r-r_h) + ... \,, \label{A-rh}\\[1mm]
&&e^{-B}=b_1 (r-r_h) + ... \,, \label{B-rh}\\[1mm]
&&\phi =\phi_h + \phi_h'(r-r_h)+ \phi'' (r-r_h)^2+ ... \,. \label{phi-rh}
\eea
The above describes a regular black-hole horizon in the presence of a scalar
field provided that $\phi'$ and the coupling function $f$ satisfy the constraints
(\ref{con-phi'})-(\ref{con-f}).



We will now show that a general coupling function $f(\phi)$ for the
scalar field does not interfere with the requirement for the existence of
an asymptotically-flat limit for the spacetime (\ref{metric}). We will
assume the following power-law expressions for the metric functions and
scalar field, in the limit $r \rightarrow \infty$,
\beq
e^{(A,B)}=1+\sum_{n=1}^\infty{\frac{(p_n,q_n)}{r}}\,,\quad 
\phi =\phi_{\infty}+\sum_{n=1}^{\infty}{\frac{d_n}{r}}\,.
\label{far}
\eeq
Upon substitution in the field equations, we may determine the arbitrary
coefficients $(p_n, q_n, d_n)$. In fact, $p_1$ and $d_1$
remain arbitrary, and we associate them with the ADM mass and scalar charge,
respectively: $p_1 \equiv -2M$ and $d_1=D$. Then, the asymptotic solutions
for the metric functions and scalar field read:
\bea
e^A&=&\; 1-\frac{2M}{r}+\frac{MD^2}{12r^3}+\frac{24MD\dot{f}+M^2D^2}{6r^4}
+...\,,\nonumber\\
e^B&=&\; 1+\frac{2M}{r}+\frac{16M^2-D^2}{4r^2}+\frac{32M^3-5MD^2}{4r^3}\nonumber\\
&+&\frac{16M(48M^3-13MD^2-24D\dot{f})+3D^4}{48r^4}+...\,,\nonumber\\
\phi&=&\; \phi_{\infty}+\frac{D}{r}+\frac{MD}{r^2}+\frac{32M^2D-D^3}{24r^3}\nonumber\\
&+&\frac{12M^3D-24M^2\dot{f}-MD^3}{6r^4}+...\,.
\eea
It is in order $\mathcal{O}(1/r^4)$ that the explicit form of the coupling function
$f(\phi)$ first makes its appearance -- indeed, the higher-curvature GB term
is expected to have a minor contribution at large distances where the curvature is
small. It is in the near-horizon regime that the GB term mainly works to support
a non-trivial scalar field, and thus a charge $D$ that then modifies the metric
compared to the Schwarzschild case.



Let us now turn our attention to the no-hair theorems, that forbid the 
existence of black-hole solutions in the presence of a scalar field, i.e.
the existence of solutions that smoothly connect the near-horizon and
far-away asymptotic solutions found above. 
We will start with the `novel' no-hair theorem developed by Bekenstein \cite{Bekenstein}.
Assuming positivity and conservation of energy, he demonstrated that the
asymptotic forms of the $T^r_{\;\,r}$ component of the energy-momentum
tensor near the horizon and at infinity could never be smoothly matched.
That argument proved beyond doubt that there were no black-hole solutions
in the context of a large class of minimally-coupled-to-gravity scalar
field theories.  Here, we will show that the coupling of the scalar field to
the quadratic GB term causes the complete evasion of Bekenstein's theorem. This may
be in fact realised for a large class of scalar field theories, with the
previously-studied exponential \cite{DBH} and linear \cite{SZ} GB couplings
comprising special cases of our present argument.

The energy-momentum tensor $T_{\mu\nu}$ satisfies the equ- ation of
conservation $D_\mu T^\mu_{\;\,\nu}=0$ due to the invariance of the action
(\ref{action}) under coordinate transformations. Its $r$-component may take
the explicit form 
\beq
(T^r_{\;\,r})'=\frac{A'}{2}\,(T^t_{\;\,t} -T^r_{\;\,r}) +
\frac{2}{r}\,(T^\theta_{\;\,\theta}-T^r_{\;\,r})\,,
\label{Trr-deriv}
\eeq
where the relation $T^\theta_\theta=T^\varphi_\varphi$ has been used due
to the spherical symmetry. The non-trivial components of the energy momentum
tensor $T_{\mu\nu}$ for our theory (\ref{action}) with a generic coupling function
$f$ are:
\bea
\hspace*{-0.5cm} 
T^t_{\;\,t}&=&-\frac{e^{-2B}}{4r^2}\left\{\phi'^2\bigl[r^2e^B+16\ddot{f}(e^B-1)\bigr]\right.
\nonumber \\ 
&-&\left.8\dot{f}\bigl[(B'\phi'(e^B-3)-2\phi''(e^B-1)\bigr]\right\},\label{Ttt}\\
T^r_{\;\,r}&=&\frac{e^{-B}\phi'}{4}\Bigl[\phi'-
\frac{8e^{-B}\left(e^B-3\right)\dot{f}A'}{r^2}\Bigr],\label{Trr}\\
\hspace*{-0.5cm}T^{\theta}_{\;\,\theta}&=&
-\frac{e^{-2B}}{4 r}\left\{\phi'^2\bigl(re^B-8\ddot{f}A'\bigr)\right. \nonumber \\
&-& \left. 4\dot{f}\bigl[\phi' (A'^2+2A'') +
A'(2\phi''-3B'\phi')\bigr]\right\}. \label{Tthth}
\eea

We will first investigate the profile of $T^r_{\;\,r}$ at infinity: using the
asymptotic expansions (\ref{far}), we easily find that
$-T^t_{\;\,t} \simeq-T^\theta_{\;\,\theta} \simeq T^r_{\;\,r} \simeq \phi'^2/4
+ {\mathcal O}\left(1/r^6\right)$.
Since the metric function $e^A$ there adopts a constant value ($e^A \rightarrow 1$),
the dominant contribution to the right-hand-side of Eq. (\ref{Trr-deriv}) is
\beq
(T^r_r)' \simeq \frac{2}{r}\,(T^\theta_\theta-T^r_r) \simeq 
-\frac{1}{r}\,\phi'^2 +....\,.
\eeq
Therefore, at asymptotic infinity, the $T^r_{\;\,r}$ component is positive and decreasing,
in agreement with \cite{Bekenstein} since the GB term is insignificant in this regime. 

In the near-horizon regime, $r \rightarrow r_h$, the $T^r_{\;\,r}$ component (\ref{Trr})
takes the approximate form:
\beq
T^r_{\;\,r}=-\frac{2e^{-B}}{r^2}A'\phi'\dot{f}+\mathcal{O}(r-r_h)\,.\label{Trr-rh}
\eeq
The above combination is finite but not negative-definite as in \cite{Bekenstein}.
In fact, the $T^r_{\;\,r}$ component {\it is} positive-definite since, close to the
black-hole horizon, $A'>0$, and $\dot f\,\phi'<0$ according to Eq. (\ref{con-phi'}). 
Therefore, the existence of a regular black-hole horizon in the context of the
theory (\ref{action}) automatically evades one of the two requirements of the
novel no-hair theorem.

Turning finally to the expression for $(T^r_r)'$ and employing the energy-momentum
components (\ref{Ttt})-(\ref{Tthth}) in Eq. (\ref{Trr-deriv}), we obtain the
following expression, in the limit $r \rightarrow r_h$:
\bea \hspace*{-0.5cm}
(T^r_r)' &=& e^{-B} A'\Bigl[-\frac{r \phi'^2}{4Z} -\frac{2 (\ddot f \phi'^2
+\dot f \phi'')}{r Z}\nonumber \\ &+&\frac{4 \dot f \phi'}{r^2}\,\Bigl(\frac{1}{r}-e^{-B} B'\Bigr)
\Bigr]+ {\mathcal O}(r-r_h)\,, 
\label{Trr'-rh}
\eea
where we have defined $Z \equiv r+2 \dot f \phi'$. Close to the black-hole
horizon, Eq. (\ref{con-phi'}) guarantees that $\dot f \phi'<0$ while $Z>0$.
Employing also the metric functions behaviour $A'>0$ and $B'<0$, we conclude
that $(T^r_r)'$ is negative in the near-horizon regime if a sole additional
constraint, namely $\ddot f \phi'^2+\dot f \phi''>0$, is satisfied. This may
be alternatively written as $\partial_r (\dot f \phi')|_{r_h}>0$, and merely
demands that the negative value of the quantity $(\dot f \phi')|_{r_h}$,
necessary for the existence of a regular black-hole solution, should be 
constrained away from the horizon. This is in fact the only way for the
matching of the two asymptotic solutions (\ref{A-rh})-(\ref{phi-rh}) and
(\ref{far}) to be realised. Therefore, for $\partial_r (\dot f \phi')|_{r_h}>0$,
the $T^r_r$ component is positive and decreasing also near the horizon
regime. As a result, both requirements of the `novel' no-hair theorem 
\cite{Bekenstein} do not apply in this theory, and thus it can be evaded.


\begin{figure}[t!]
\minipage{0.46\textwidth}
  \includegraphics[width=\linewidth]{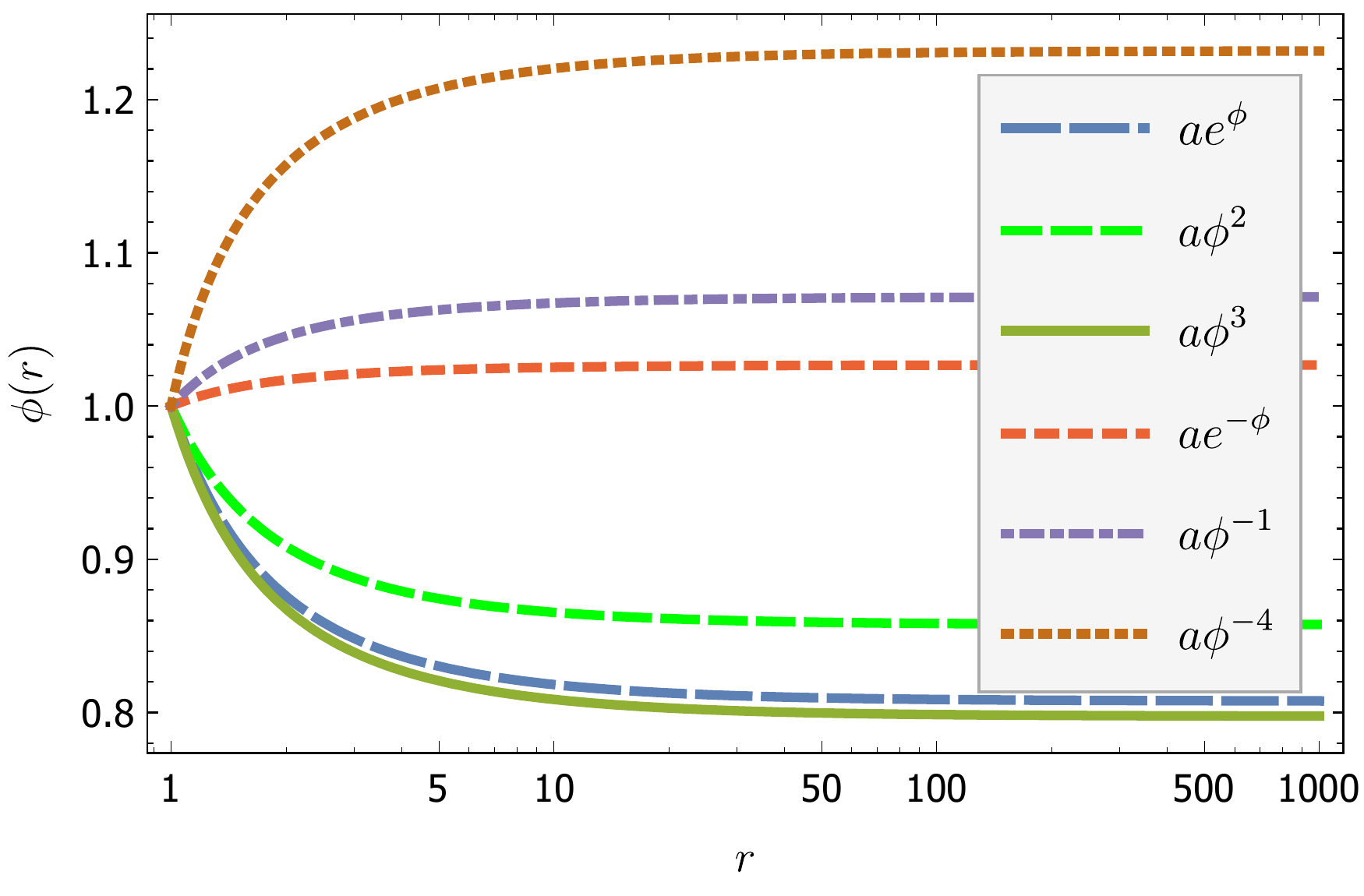}
  \caption{The scalar field $\phi$ for different coupling functions $f(\phi)$,
  for $a=0.01$ and $\phi_h=1$.} \label{Fig_phi}
\endminipage\hfill
\end{figure}

The older version of the no-hair theorem for scalar fields \cite{NH-scalar},
that employs the scalar equation of motion, also fails to exclude the existence of
black-hole solutions in our theory (\ref{action}): multiplying the scalar equation
(\ref{scalar-eq}) by $f(\phi)$ and integrating over the black-hole exterior region, we
obtain the integral constraint
\beq
\int d^4x \sqrt{-g} \,f(\phi) \left[\nabla^2 \phi + \dot f(\phi) R^2_{GB} \right] =0\,.
\eeq
Integrating by parts the first term, the above becomes
\beq
\int  d^4x \sqrt{-g}\,\dot f(\phi) \left[\partial_\mu \phi\,\partial^\mu \phi  - 
f(\phi)  R^2_{GB}\right] =0\,. \label{old-con}
\eeq
The boundary term $[\sqrt{-g} f(\phi) \partial^\mu \phi]_{r_h}^\infty$
vanishes both at the horizon (due to the $e^{(A-B)/2}$ factor)
and at infinity (due to the  $\partial^\mu \phi$ factor). Since $\phi=\phi(r)$,
the first term in Eq. (\ref{old-con}) gives
$\partial_\mu \phi\,\partial^\mu \phi = g^{rr}(\partial_r \phi)^2>0$ 
throughout the exterior region.  Also, for the metric (\ref{metric}),
the GB term has the explicit form
\beq
R^2_{GB} =\frac{2 e^{-2B}}{r^2}\bigl[(e^B-3)\,A' B' -(e^B-1)\,(2A''+A'^2)
\bigr]. \label{GB}
\eeq
Employing the asymptotic solutions near the horizon (\ref{A-rh})-(\ref{B-rh})
and at infinity (\ref{far}), we may easily see that the GB term
takes on a positive value at both regimes. Therefore, in the simplest
possible case where both $f(\phi)$ and $R^2_{GB}$ are sign-definite,
Eq. (\ref{old-con}) allows for black-hole solutions with scalar hair
for every choice of the coupling function that merely satisfies $f(\phi)>0$.


\begin{figure}[t!]
\minipage{0.46\textwidth}
  \includegraphics[width=\linewidth]{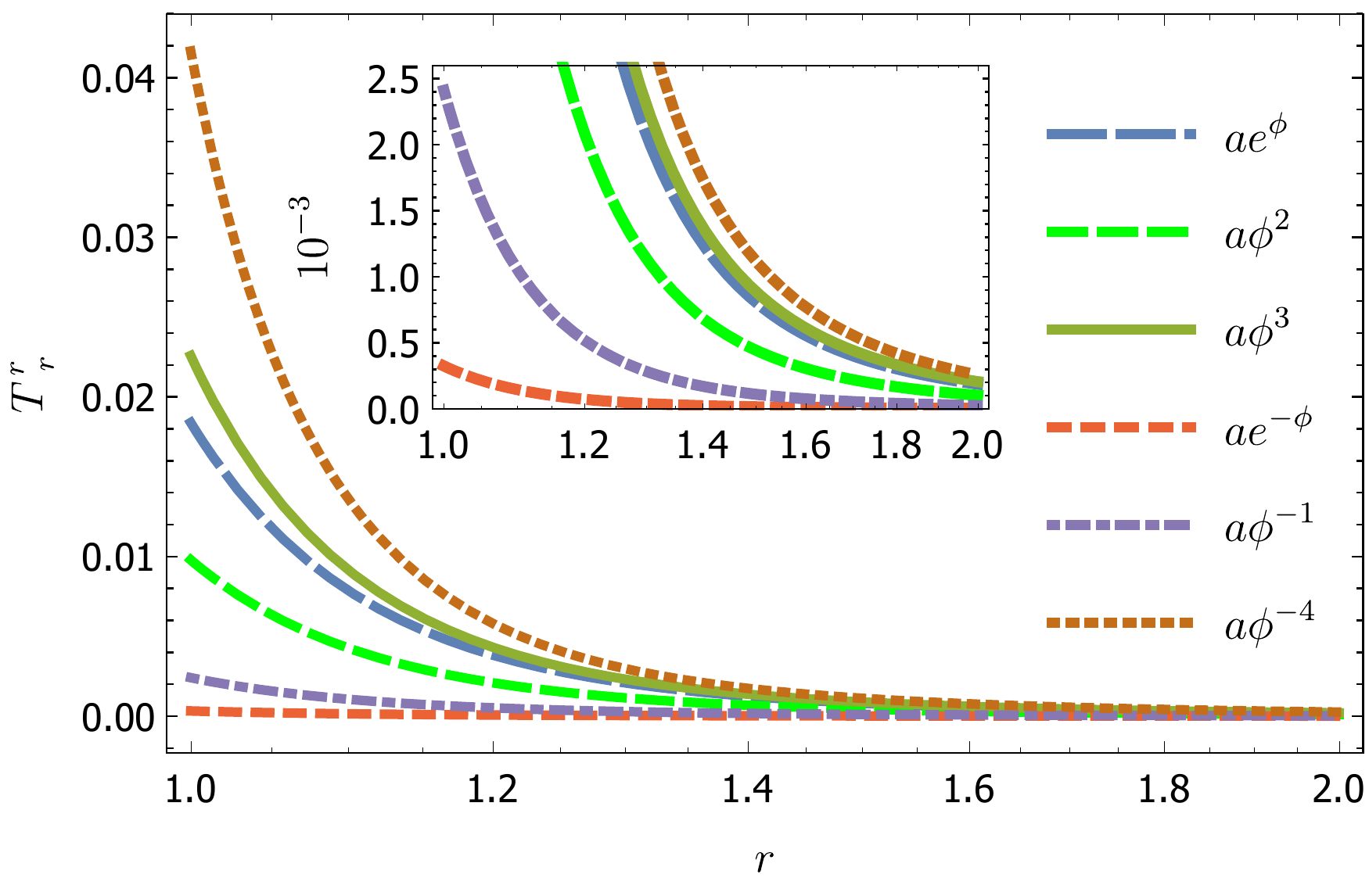}
  \caption{The $T^r_{\,\,r}$ component for different coupling functions $f(\phi)$,
  for $a=0.01$ and $\phi_h=1$.} \label{Fig_emt}
\endminipage\hfill
\end{figure}


In order to demonstrate the validity of the aforementioned arguments, we
have numerically solved the system of equations (\ref{Aphi-system}), and
produced a large number of black-hole solutions with scalar hair. The solutions
for the scalar field are depicted in Fig. \ref{Fig_phi}, for a variety of
forms of the coupling function $f(\phi)$: exponential, odd and even power-law,
odd and even inverse-power-law. These forms are all simple, natural choices to
keep the GB term in the 4-dimensional theory. For easy comparison, the coupling
constant in all cases has been set to $\alpha =0.01$ and the near-horizon value
of the field to $\phi_h=1$. For $f(\phi)=(\alpha e^\phi, \alpha \phi^2, 
\alpha \phi^3)$, that all have $f'_h>0$, our constraint (\ref{con-phi'}) leads
to a negative $\phi'_h$; for
$f(\phi)=(\alpha e^{-\phi}, \alpha \phi^{-1}, \alpha \phi^{-4})$, that
have $f'_h<0$, Eq. (\ref{con-phi'}) demands a positive $\phi'_h$ -- the
decreasing and increasing, respectively, profiles are clearly depicted
in Fig. \ref{Fig_phi}. In all cases, for a given value of $\phi_h$,
Eq. (\ref{con-phi'}) uniquely determines the quantity $\phi'_h$.
The integration of the system (\ref{Aphi-system}) with initial conditions
$(\phi_h, \phi'_h)$ then leads to the presented solutions.
The positivity and decreasing profile of the $T^r_{\,\,r}$ component, 
necessary features for the evasion of the novel no-hair theorem, are clearly
seen in Fig. \ref{Fig_emt}. It is worth mentioning that the second constraint,
$\partial_r (\dot f \phi')|_{r_h}>0$, is automatically satisfied for every
solution found and does not need any further action or fine-tuning of the
free parameters. We finally note that, for $\phi>0$, all the above forms
of $f(\phi)$ satisfy also the constraint $f(\phi)>0$, derived above for
the evasion of the old no-hair theorem.

For a given coupling function $f(\phi)$, and fixed ($\alpha$, $\phi_h$),
the constraint (\ref{con-f}) dictates that there is a lower bound for the
horizon radius of the derived black-hole solutions given by
$r_h^2 > 4\sqrt{6}\,|\dot f_h|$, and thus a lower-bound on their mass.
This characteristic has been noted in the exponential-coupling
case \cite{DBH, Blazquez}, and comprises a generic feature of all the GB
black holes found here, that distinguishes them from their GR analogues. 
Thus, in the small-mass limit, observable effects may include deviations
from standard GR in the calculation of the bending angle
of light, the precession observed in near-horizon orbits and the spectrum
from their accretion discs \cite{Chakra}. The emission of scalar radiation
strongly depends to the existing coupling of the scalar field to ordinary
matter, while the measurement of the characteristic frequencies
of the quasi-normal modes (especially the polar sector) will also help to
distinguish these solutions from their GR analogues \cite{Kunz}. 
Finally, the detection of gravitational waves from black-hole or neutron-star
mergers may also help to impose bounds on the parameters of the theory provided
that the scalar charge is significant and their physical distance is small.


\section{The Scalar-GB Theory}

Finally, we investigate whether a regular black-hole solution can arise
as the result of the synergy between only the scalar field $\phi$ and the GB term.
To this end, we ignore all terms in the field equations coming from the Ricci
scalar, and attempt to construct a near-horizon, regular solution similar to that
of Eqs. (\ref{A-rh})-(\ref{phi-rh}). In the absence of all $R$-related terms in
the field equations, the components of $T_{\mu\nu}$ should vanish. If we assume
again that, as $r \rightarrow r_h$, $\phi'$ remains finite while $A'$ diverges,
Eq. (\ref{Trr}) yields $e^B \simeq 3 +{\cal O}(1/A')$; this clearly does not
describe a black hole. Alternatively, demanding that $e^B \rightarrow \infty$,
as $r \rightarrow r_h$, Eq. (\ref{Trr}) may be solved for $A'$ to give:
$A' \simeq r^2\phi'/8\dot{f}+\mathcal{O}(e^{-B})$. 
Using this, Eqs. (\ref{Ttt}) and (\ref{Tthth}) form a system of two differential
equations for $B$ and $\phi$. In the limit $r \rightarrow r_h$, we find
the behaviour
\bea
B'&=&-\frac{2}{r}\,e^B+\mathcal{O}\left(e^{-B}\right),\label{106}\\
\phi''&=&-\frac{e^B}{r}\,\phi'+\mathcal{O}\left(e^{-B}\right).\label{107}
\eea
Equation (\ref{106}) leads to the solution: $e^{-B}=2 \ln\left(r/r_h\right)$,
which does point towards the existence of a horizon. However, for this
horizon to be regular, Eq. (\ref{107}) demands that $\phi'(r_h)=0$. But
then, $\phi''(r_h)$ is also zero leading to a constant scalar field outside
the horizon. In this case, the GB term does not contribute to the field
equations and the above solution disappears.


\section{Conclusions}

In the context of a general Einstein-scalar-GB theory with an arbitrary coupling
function $f(\phi)$, we have demonstrated that the emergence of regular black-hole
solutions is a generic feature: the explicit
form of $f(\phi)$ affects very little the asymptotically-flat limit at infinity,
while a regular horizon is formed provided that $\phi'_h$ and $f(\phi)$ satisfy
the constraints (\ref{con-phi'})-(\ref{con-f}). 

The existing no-hair theorems were shown to be evaded under mild assumptions
on $f(\phi)$. The old no-hair theorem \cite{NH-scalar} is easily evaded for 
$f(\phi)>0$ while the novel no-hair theorem \cite{Bekenstein} is non-applicable
if the same constraint (\ref{con-phi'}) holds. Based on this, we have produced
a large number of regular black-hole solutions with non-trivial scalar hair for
arbitrary forms of the coupling function $f(\phi)$. They are all characterised
by a minimum black-hole radius and mass, and their near-horizon strong dynamics
is expected to leave its imprint on a number of observables. The obtained
solutions survive only when the synergy of $\phi$ with the GB term is 
supplemented by the linear Ricci term.


{\bf Aknowledgement} P.K. is grateful to Naresh Dadhich for a constructive
communication at the early stages of this work.

{\bf Note added:} After our paper was submitted for publication, two
additional works appeared \cite{Doneva, Silva} where particular solutions
of our general theory are discussed.


\begin{thebibliography}{9}

\bibitem{NH-scalar} J.~D.~Bekenstein,
  Phys.\ Rev.\ Lett.\  {\bf 28} (1972) 452; 
C.~Teitelboim,
  Lett.\ Nuovo Cim.\  {\bf 3S2} (1972) 397.

\bibitem{YM} M.~S.~Volkov and D.~V.~Galtsov,
 JETP Lett.\  {\bf 50} (1989) 346;
P.~Bizon,
  Phys.\ Rev.\ Lett.\  {\bf 64} (1990) 2844;
B.~R.~Greene, S.~D.~Mathur and C.~M.~O'Neill,
  Phys.\ Rev.\ D {\bf 47} (1993) 2242;
K.~I.~Maeda, T.~Tachizawa, T.~Torii and T.~Maki,
Phys.\ Rev.\ Lett.\  {\bf 72} (1994) 450.


\bibitem{Skyrmions} H.~Luckock and I.~Moss,
  Phys.\ Lett.\ B {\bf 176} (1986) 341;
S.~Droz, M.~Heusler and N.~Straumann,
  Phys.\ Lett.\ B {\bf 268} (1991) 371.

\bibitem{Conformal} J.~D.~Bekenstein,
  Annals Phys.\  {\bf 82} (1974) 535; 
Annals Phys.\  {\bf 91} (1975) 75.

\bibitem{Bekenstein} J.~D.~Bekenstein,
  Phys.\ Rev.\ D {\bf 51} (1995) no.12,  R6608.
  
\bibitem{Herdeiro}
  C.~A.~R.~Herdeiro and E.~Radu,
  Int.\ J.\ Mod.\ Phys.\ D {\bf 24} (2015) no.09,  1542014.

\bibitem{SF} T.~P.~Sotiriou and V.~Faraoni,
  Phys.\ Rev.\ Lett.\  {\bf 108} (2012) 081103.

\bibitem{HN}  L.~Hui and A.~Nicolis,
  Phys.\ Rev.\ Lett.\  {\bf 110} (2013) 241104.

\bibitem{DBH} P.~Kanti, N.~E.~Mavromatos, J.~Rizos, K.~Tamvakis and E.~Winstanley,
  Phys.\ Rev.\ D {\bf 54} (1996) 5049; 
Phys.\ Rev.\ D {\bf 57} (1998) 6255.
  
\bibitem{Babichev}
  E.~Babichev and C.~Charmousis,
 JHEP {\bf 1408} (2014) 106.

\bibitem{SZ}  T.~P.~Sotiriou and S.~Y.~Zhou,
  Phys.\ Rev.\ Lett.\  {\bf 112} (2014) 251102.

\bibitem{Benkel}  T.~P.~Sotiriou and S.~Y.~Zhou,
  Phys.\ Rev.\ D {\bf 90} (2014) 124063;
R.~Benkel, T.~P.~Sotiriou and H.~Witek,
  Class.\ Quant.\ Grav.\  {\bf 34} (2017) no.6,  064001;
Phys.\ Rev.\ D {\bf 94} (2016) no.12,  121503.


\bibitem{Metsaev} R.~R.~Metsaev and A.~A.~Tseytlin,
  Nucl.\ Phys.\ B {\bf 293} (1987) 385.
  
\bibitem{Horndeski} G.~W.~Horndeski,
  Int.\ J.\ Theor.\ Phys.\  {\bf 10} (1974) 363.
  
  
\bibitem{Blazquez} J.~L.~Blazquez-Salcedo {\it et al.},
  IAU Symp.\  {\bf 324} (2016) 265 [arXiv:1610.09214 [gr-qc]].
  
  \bibitem{Chakra} S.~Bhattacharya and S.~Chakraborty,
  Phys.\ Rev.\ D {\bf 95} (2017) no.4,  044037;
  I.~Banerjee, S.~Chakraborty and S.~SenGupta,
  Phys.\ Rev.\ D {\bf 96} (2017) no.8,  084035.

\bibitem{Kunz} 
J.~L.~Blazquez-Salcedo, C.~F.~B.~Macedo, V.~Cardoso, V.~Ferrari, L.~Gualtieri, F.~S.~Khoo, J.~Kunz and P.~Pani,
  Phys.\ Rev.\ D {\bf 94} (2016) no.10,  104024.
  
\bibitem{Doneva} D.~D.~Doneva and S.~S.~Yazadjiev,
  arXiv:1711.01187 [gr-qc].

\bibitem{Silva} H.~O.~Silva, J.~Sakstein, L.~Gualtieri, T.~P.~Sotiriou and E.~Berti,
  arXiv:1711.02080 [gr-qc].



\end{thebibliography}
\end{document}